\documentclass[twocolumn,amsmath,amssymb,floatfix,prl,shownopacs,footinbib,superscriptaddress]{revtex4-1}
\usepackage{amsmath,amssymb,natbib,bm,graphicx,url,epsf,color}
\usepackage[british]{babel}
\usepackage{pdfpages}
\usepackage{wasysym}
\usepackage{MnSymbol}

\newcommand{\Vdc}{V_{\mathrm{dc}}}
\newcommand{\Vac}{V_{\mathrm{ac}}}
\newcommand{\Vem}{V_{\mathrm{em}}}

\begin{document}

\title{Photon-assisted shot noise in graphene in the Terahertz range}

\author{F.D. Parmentier}
\affiliation{SPEC, CEA, CNRS, Universit\'e Paris-Saclay, CEA Saclay 91191 Gif-sur-Yvette cedex, France
}
\author{L.N. Serkovic-Loli}
\affiliation{SPEC, CEA, CNRS, Universit\'e Paris-Saclay, CEA Saclay 91191 Gif-sur-Yvette cedex, France
}\affiliation{Instituto de F\'isica, Universidad Nacional Aut\'onoma de M\'exico, Coyoacan 04510 Ciudad de M\'exico, M\'exico}
\author{P. Roulleau}
\affiliation{SPEC, CEA, CNRS, Universit\'e Paris-Saclay, CEA Saclay 91191 Gif-sur-Yvette cedex, France
}
\author{D.C. Glattli}
\affiliation{SPEC, CEA, CNRS, Universit\'e Paris-Saclay, CEA Saclay 91191 Gif-sur-Yvette cedex, France
}

\date{\today}

\begin{abstract}
When subjected to electromagnetic radiation, the fluctuation of the electronic current across a quantum conductor increases. This additional noise, called photon-assisted shot noise, arises from the generation and subsequent partition of electron-hole pairs in the conductor. The physics of photon-assisted shot noise has been thoroughly investigated at microwave frequencies up to 20 GHz, and its robustness suggests that it could be extended to the Terahertz (THz) range. Here, we present measurements of the quantum shot noise generated in a graphene nanoribbon subjected to a THz radiation. Our results show signatures of photon-assisted shot noise, further demonstrating that hallmark time-dependant quantum transport phenomena can be transposed to the THz range.

\end{abstract}

\maketitle

The many promises of the Terahertz (THz) frequency range, in terms of both fundamental and practical applications, has led to the increasingly active development of various THz sources and detectors~\cite{Ferguson2002,Tonouchi2007,Rogalski2011}. This recently allowed the study of fundamental aspects of time-dependent electronic quantum transport at higher frequencies, comparable with the characteristic energy scales arising in highly confined electronic systems, such as carbon nanotubes, self-assembled semiconductor quantum dots, and single molecule transistors~\cite{Kawano2008,Shibata2012,Yoshida2015}. These previous works focused on photon-assisted tunnelling (PAT), wherein electron transport across a discrete electronic level is mediated by the absorption of a resonant impinging photon~\cite{Tien1963,Tucker1985,Kouwenhoven1994}. The ability to reliably couple THz-range radiation to electronic transport degrees of freedom in a quantum conductor significantly broadens the range of exploration of the influence of electron-photon interactions on charge transport. These interactions, which have been extensively studied in the microwave range, can give rise to striking modifications of the conductance of a coherent conductor, by either increasing it, as it is the case for PAT, or strongly suppressing it in the so-called dynamical Coulomb blockade~\cite{Nazarov1992}.

Electron-photon interactions can have subtle effects, which do not appear in the electronic conductance, but rather in the fluctuations of the current across the quantum conductor, or quantum shot noise. Such is the case for photon-assisted shot noise (PASN), a hallmark of time-dependent electronic quantum transport where incident photons excite electron-hole pairs in the leads of a coherent conductor~\cite{Lesovik1994,Blanter2000,Schoelkopf1998,Kozhevnikov2000,Reydellet2003,Gasse2013a,Gabelli2013,Dubois2013,Jompol2015}. The electron-hole pairs then propagate in the conductor, in which quantum partitioning leads to an increase of shot noise, while the net current remains zero. This increase can be expressed as an equivalent noise temperature $T_N$:
\begin{multline}
 T_N=  T_{el} +  F \sum_{n} J_n^2(\frac{e \Vac}{h\nu})\\
 \times\frac{e \Vdc+n h\nu}{2 k_B}\left[\mathrm{coth}\left(\frac{e \Vdc+n h\nu}{2 k_B T_{el}}\right)-\frac{2 k_B T_{el}}{e \Vdc+n h\nu}\right],
\label{eq-PASN}
\end{multline}
where $\Vdc$ is the dc drain-source voltage applied to the conductor, $V_{ac}$ and $\nu$ are the amplitude and frequency of the electromagnetic radiation, $T_{el}$ is the electron temperature, $F$ is the Fano factor characterizing transport in the conductor, $J_n$ are Bessel functions of the first kind, $e$, $h$ and $k_B$ are respectively the electron charge, Planck's and Boltzmann constants.
 Predicted more than two decades ago~\cite{Lesovik1994}, and extensively studied in the microwave domain~\cite{Schoelkopf1998,Kozhevnikov2000,Reydellet2003,Gasse2013a,Gabelli2013,Dubois2013,Jompol2015}, PASN remarkably allows reconstructing the out-of equilibrium energy distribution function arising in the conductor due to the time-dependent potential~\cite{Gabelli2013,Dubois2013}, as well as calibrating the amplitude and frequency of a monochromatic radiation impinging on the conductor~\cite{Gasse2013a}.

\begin{figure}[h!]
\centering\includegraphics[width=0.8\columnwidth]{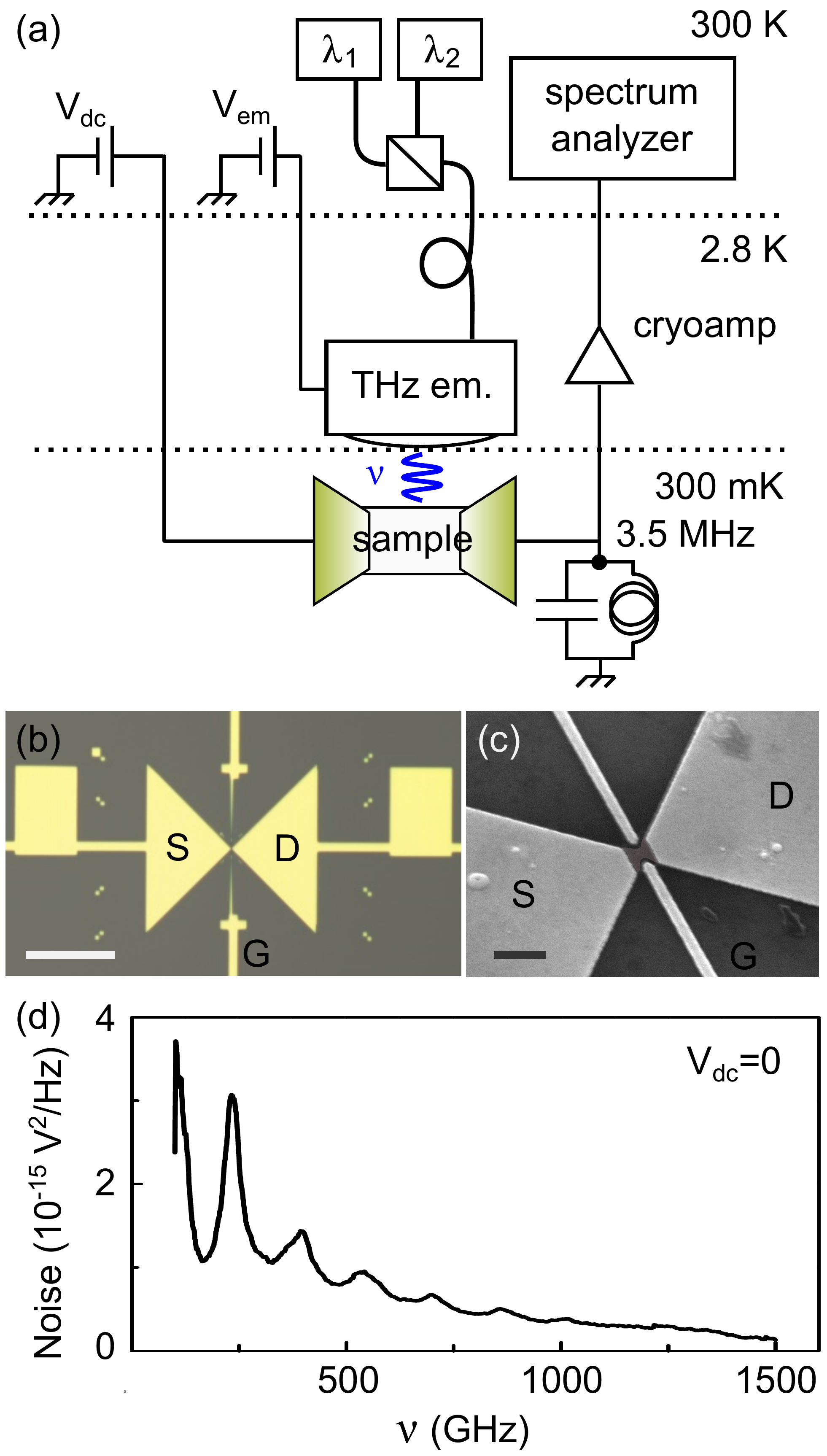}
\caption{
(a), Simplified schematic description of our setup combining a photomixing THz source and noise measurements. The THz emitter (thermally anchored to the 2.8 K stage of the refrigerator) is aligned in front of the sample (set at 300 mK), the shot noise of which is measured through a resonant LC circuit. (b), Optical micrograph of a typical sample, showing the bow-tie antenna-shaped electrodes (\emph{S} and \emph{D}). The scale bar corresponds to 50~$\mu$m. (c), Scanning electron micrograph of a typical sample, showing the tip of the electrodes, as well as the side gate (\emph{G}). The shape of the graphene ribbon is shown in light grey. The scale bar corresponds to 1~$\mu$m. (d), Integrated excess noise power at the output of the setup as a function of the frequency of the THz radiation, measured on sample A.
}
\label{fig-fig1}
\end{figure}

To unveil the signatures of PASN due to THz radiation, we have measured the shot noise of graphene coherent conductors in presence of a THz excitation. Graphene, which has been shown to host a variety of striking linear and nonlinear optical effects, such as wide-spectrum saturable absorption~\cite{Bonaccorso2010}, is particularly well suited for THz applications, its high mobility~\cite{Mayorov2011} and low electron-phonon coupling~\cite{Balandin2011} allowing its use in a large number of THz detectors based on different mechanisms~\cite{Koppens2014}. We rely on the ability to easily engineer ribbons of disordered graphene, in which electronic transport is diffusive~\cite{Chen2009}. Using a diffusive conductor has several advantages for the noise measurements presented here. First, the sample conductance is essentially independent of the energy up to high energies. This ensures that, in absence of THz excitation, the shot noise is indeed linear with the drain-source voltage $\Vdc$, and devoid of features which would mask the signatures of PASN. This also allows neglecting the effects of photon-assisted tunnelling (which only occurs in energy-dependent conductors), which would generate an additional shot noise with strong dependences in $\Vdc$, again potentially masking the signatures of PASN. Second, the value of the Fano factor $F=1/3$ is well known for diffusive conductors~\cite{Blanter2000}, simplifying the analysis.

Fig.~\ref{fig-fig1}(a) shows a simplified description of our experimental setup~\cite{SM}. We use a Toptica cw THz generator based on a photomixing technique~\cite{Deninger2008,Roggenbuck2010} to illuminate the samples in the 50 GHz - 2 THz range. The emitter, consisting of a rapid photo-switch coupled to a focusing Silicon lens, is aligned in vacuum a few mm above the sample, which is cooled down to 300 mK. The (uncalibrated) power of the THz radiation is modulated by the bias voltage $\Vem$ applied to the diode. Fig.~\ref{fig-fig1}(b) and~\ref{fig-fig1}(c) show optical and scanning electron micrographs of a typical sample, consisting of a CVD grown monolayer graphene nanoribbon connected to source and drain electrodes shaped as the two parts of a bow-tie antenna. Side gates (labelled $G$) allows tuning the transport through the nanoribbon. A dc voltage $\Vdc$ is applied across the sample, and the power spectral density of the shot noise it generates is filtered through an LC tank with a resonance frequency $f_{\mathrm{LC}}\approx3.5$~MHz, and a bandwidth at half maximum $\Delta f_{\mathrm{LC}}\approx0.4$~MHz. This is crucial to increase the sensitivity and stability of the measurement, as $1/f$ noise and microphonics are absent in this frequency range~\cite{Dicarlo2006}. The noise signal is then amplified using home-made cryogenic amplifiers (input voltage noise $S_{\mathrm{V,}amp}\approx0.14~\mathrm{nV}/\sqrt{\mathrm{Hz}}$).
Fig.~\ref{fig-fig1}(d) shows the uncalibrated noise signal at the end of our detection chain as a function of the frequency $\nu$ of the THz radiation at maximum power ($\Vem=13$~V), for a first sample (labelled A) at $\Vdc=0$. Clear oscillations appear as $\nu$ is swept; however, the precise frequency dependence of the signal is challenging to analyze, as it is not reproducible from sample to sample~\cite{SM}, and contains contributions of the emitter power frequency dependence (which is monotonously decreasing, with about -40 dB$/$decade~\cite{Deninger2008,Roggenbuck2010}), standing waves between the emitter and the samples, and of the frequency response of the sample's antennas. A simple numerical simulation of the frequency response of the antenna showed resonances qualitatively similar to Fig.~\ref{fig-fig1}(d)~\cite{SM}. This nonetheless confirms that noise measurements in graphene and other nano-devices can be used for THz detection~\cite{Santavicca2010,Wang2014}.

\begin{figure}[h!]
\centering\includegraphics[width=0.9\columnwidth]{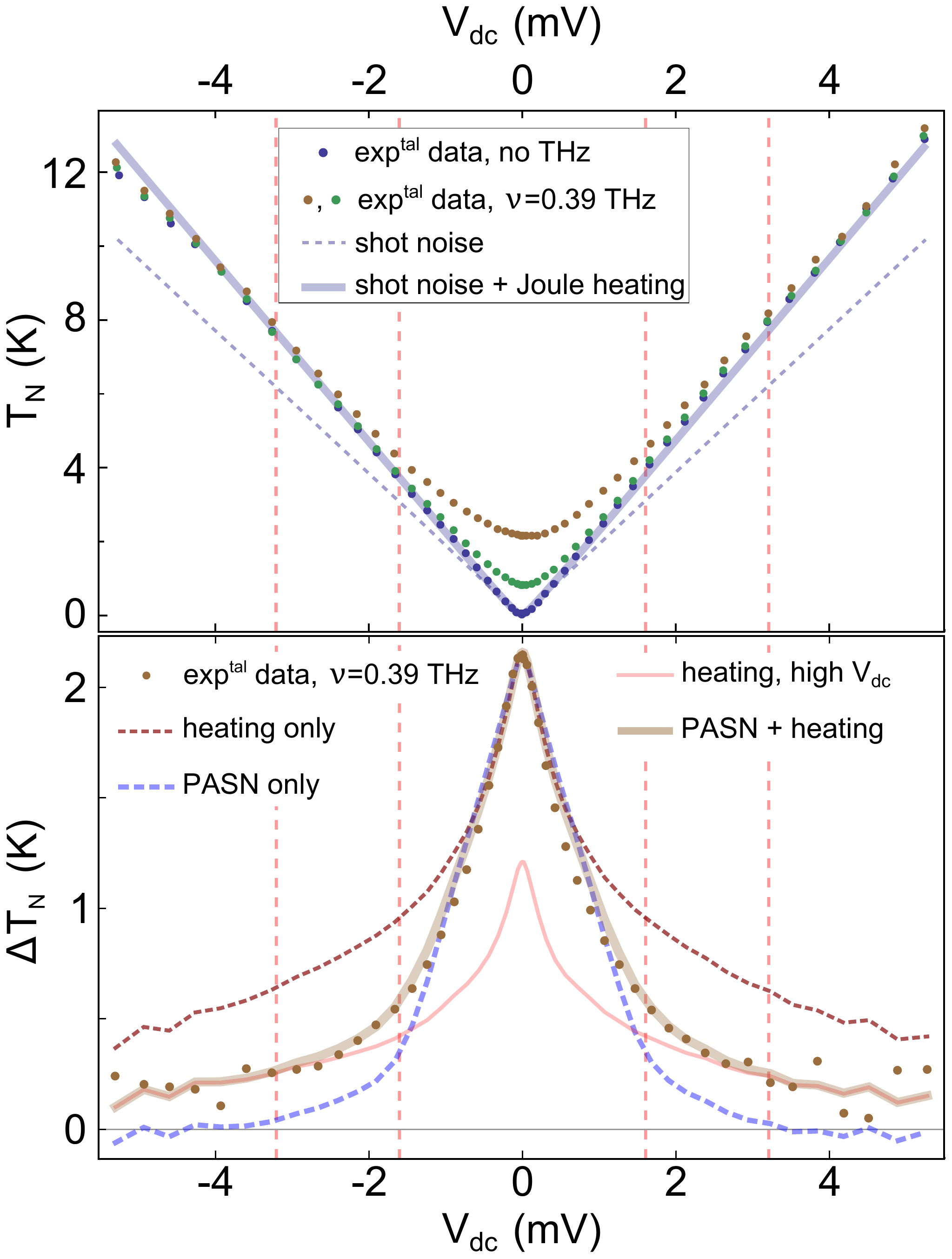}
\caption{
 Upper panel: excess noise temperature $T_N$ as a function of the dc bias applied to the sample (sample B). The purple, green and brown symbols are the experimental data, respectively in absence of THz radiation ($\Vem=0$~V), at $\nu=0.39$~THz and $\Vem=10$~V, and at $\nu=0.39$~THz and $\Vem=13$~V. The dashed line is the expected shot noise in absence of heating, and the continuous line is a fit of the data without THz radiation, including heating. Lower panel: difference $\Delta T_N$ of the noise in presence ($\Vem=13$~V) and in absence of THz radiation. The red and blue dashed lines are fits of the experimental data using, respectively, a heating model and a PASN model. The thin continuous line is a fit of the data at large $\Vdc$ using a heating model, and the thick line is a fit combining this last model and a PASN model at low $\Vdc$. In both panels, the symbols size corresponds to the statistical error on the measurement. The red vertical dashed lines in both panels correspond to $e\Vdc=\left\{-2,-1,1,2\right\}h\nu$.
}
\label{fig-fig2}
\end{figure}

We now analyze the origin of the noise increase by measuring its dependence with $\Vdc$ in a second sample, labelled B. The upper panel of Fig.~\ref{fig-fig2} shows the measured variations of the calibrated noise temperature $T_N=S_i / 4 k_B G_s$ with $\Vdc$, where $S_i$ is the current noise generated in the sample and $G_s$ is the conductance of the sample, measured simultaneously using standard lock-in techniques~\cite{SM}. We first show that in absence of THz radiation ($\Vem=0$, purple symbols), significant heating effects arise due to Joule power dissipated in the leads: the dashed line in the upper panel of Fig.~\ref{fig-fig2}, corresponding to shot noise with the Fano factor $F=1/3$ of diffusive coherent conductors and no heating, is markedly lower than the data. Importantly, the linear variation of $T_N$ with $\Vdc$ implies that cooling is only mediated by the electronic transport channels via the Wiedemann-Franz law, where the cooling power is proportional to the temperature $T_{el}$ squared~\cite{SM}. Indeed, if cooling mediated by electron-phonon coupling were important in the system, the noise would present sub-linear features, the cooling power due to electron-phonon coupling in graphene being proportional to $T_{el}^\delta$, with $\delta$ typically equal to 4~\cite{Betz2012a}. A model using only the Wiedemann-Franz law yields the continuous line, in excellent agreement with the data. The ratio between the resistance of the graphene nanoribbon and the contact resistance is used as the fit parameter~\cite{SM,Kumar1996}. 
When the THz radiation at $\nu=0.39$~THz is turned on (green and brown symbols, for an emitter voltage $\Vem$ of resp. 10 V and 13 V), the noise clearly increases at low $\Vdc$, then approaches the data without THz radiation at $\left|\Vdc\right|>2$~mV. The increase at low $\Vdc$ is directly related to the voltage $\Vem$, and thus to the power of the THz radiation. The effect of the radiation appears more clearly when plotting the difference $\Delta T_N$ between the noise in presence and in absence of radiation, as shown in the lower panel of Fig.~\ref{fig-fig2}, for $\nu=0.39$~THz and $\Vem=13$~V. In particular, $\Delta T_N$ sharply decreases as $\left|\Vdc\right|$ increases on a typical energy scale given by $e\left|\Vdc\right|=h\nu$ (red dashed vertical lines), then saturates to a finite value at large $\left|\Vdc\right|$. This behaviour cannot be quantitatively reproduced by models describing the effect of THz as either pure PASN, or simple heating. Since the THz power impinging on the sample is not known by construction of the experiment, we use it (either as an ac amplitude $\Vac$, or a temperature increase $T_{\mathrm{THz}}$) as the fit parameter in these models. In the first model (corresponding to the blue dashed line in the lower panel of Fig.~\ref{fig-fig2}), the noise is given by the PASN described in Eq.~\ref{eq-PASN}, where $\alpha=e\Vac/h\nu$ is adjusted to fit the data at $\Vdc=0$, yielding $\alpha=1.3$, and the electronic temperature $T_{el}(\Vdc)$ is extracted from the fit of the shot noise in absence of THz. Notably, the predicted $\Delta T_N$ is zero for $\left|e\Vdc\right|>2h\nu$, as PASN reduces to the usual shot noise when $\Vdc\gg\Vac$, whereas the experimental data remains finite. In the second model (red dashed line), the noise is given by the usual shot noise expression (\textit{i.e.} without ac excitation)~\cite{Blanter2000}, where the electronic temperature is increased by a constant amount $T_{\mathrm{THz}}$, adjusted to fit the data at $\Vdc=0$: $T_{el}^{\ast}(\Vdc)=\sqrt{T_{el}(\Vdc)^2+T_{\mathrm{THz}}^2}$, with $T_{\mathrm{THz}}=2.41$~K. This dependence again stems from the fact that only electronic channels contribute to heat transport in the sample. Because of this dependence, the predicted noise decreases more slowly than the predicted PASN, or indeed, our experimental data. Note that 1) the statistical error on our data is much smaller than the difference between our data and either model, and that 2) regardless of the fitting procedure (adjusting the large $\left|\Vdc\right|$ value of the noise, or the entire curve), neither model allow to accurately describe our data. Since the experimental data clearly sits in between the results of both models, particularly at large $\left|\Vdc\right|$, we interpret it using a model combining both heating and PASN. We first extract $T_{\mathrm{THz}}=1.45$~K using the heating model to fit the data at large $\Vdc$ (thin continuous line), where the PASN theory predicts zero excess noise, then insert the increased temperature in the PASN formula while adjusting $\alpha=0.89$ to match the data at $\Vdc=0$. The result of this model, shown as a thick continuous line, is in excellent agreement with the data, over the whole range of explored $\Vdc$.

\begin{figure}[h!]
\centering\includegraphics[width=0.9\columnwidth]{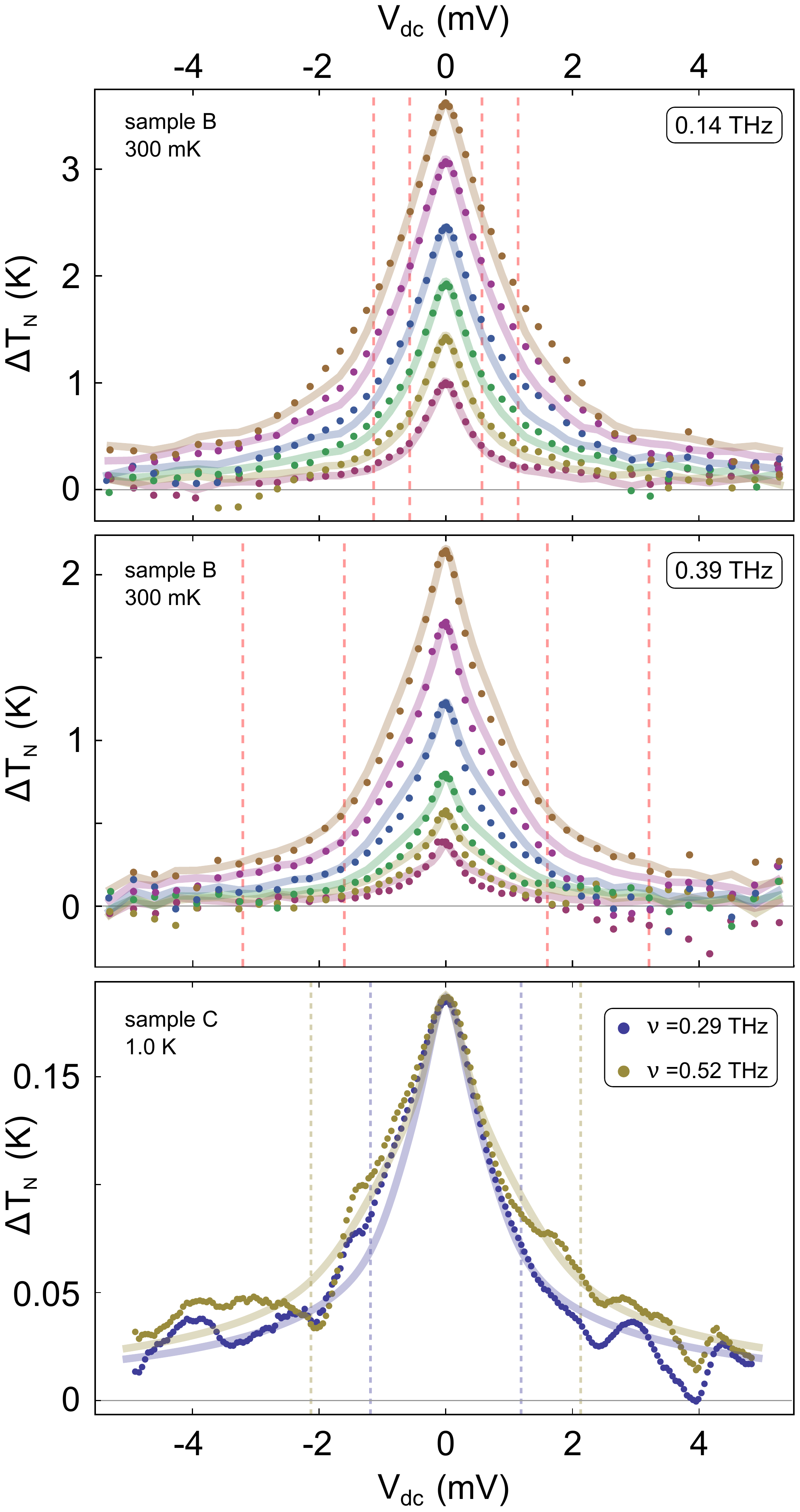}
\caption{
Top and central panels: noise difference $\Delta T_N$ for $\nu =$0.14 and 0.39 THz, measured at 300 mK on sample B. In both, $\Vem$ is changed from 8 (dark purple) to 13 V (dark yellow). Bottom panel: $\Delta T_N$ measured at 1 K on sample C for $\nu =$0.29 (blue) and 0.52 THz (dark yellow) and $\Vem=$13 V. In all panels, symbols are the experimental data, and continuous lines are fits combining heating and PASN, as explained in Fig.~\ref{fig-fig2}.
}
\label{fig-fig3}
\end{figure}

Fig.~\ref{fig-fig3} shows the application of this analysis on experimental data obtained for various THz frequencies ($\nu$ ranging from 0.14 to 0.52 THz) and powers ($\Vem$ ranging from 8 to 13 V), and two different samples (sample B and C, respectively measured at 300 mK and 1 K). The excellent general agreement confirms the validity of our interpretation. While hot-electron effects smear the structures at $e\Vdc=h\nu$ expected from the PASN theory, the influence of the THz frequency on the noise can be seen as a broadening of the noise difference $\Delta T_N$ as a function of $\Vdc$ as $\nu$ increases, as shown in the lower panel of Fig.~\ref{fig-fig3}. 

\begin{figure}[!htbp]
\centering\includegraphics[width=0.9\columnwidth]{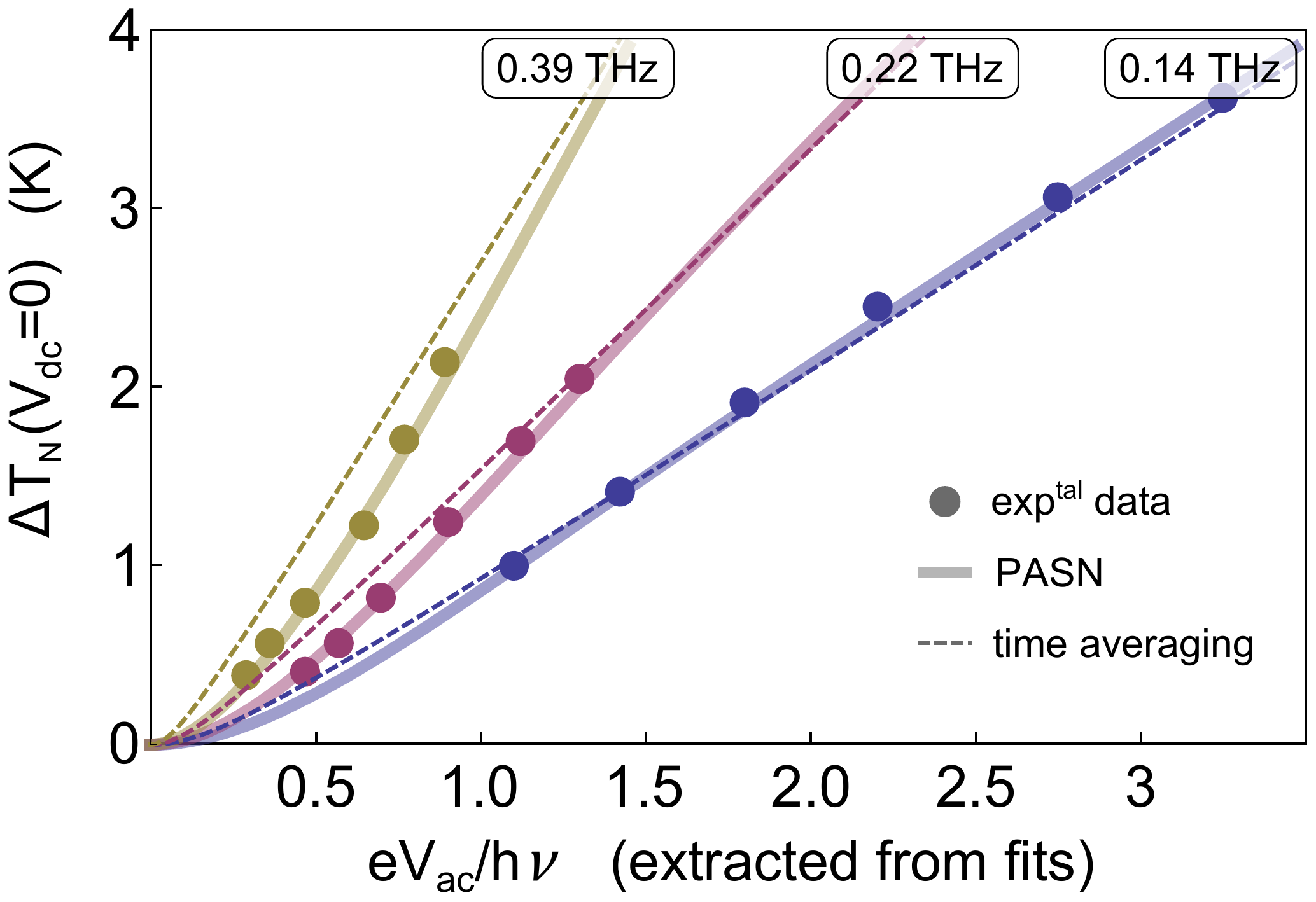}
\caption{
Noise difference $\Delta T_N$ at $\Vdc=0$ measured on sample B for $\nu =$0.14, 0.22 and 0.39 THz, as a function of $\alpha=e\Vac/h\nu$ extracted from the fits explained in Fig.~\ref{fig-fig2}. The continuous lines are predictions of the model combining PASN and heating, and the dashed lines are predictions of the model combining heating and time-averaging of the shot noise. The size of the symbols indicates the uncertainty on the extraction of $\alpha$.}
\label{fig-fig4}
\end{figure}

To emphasize the effect of $\nu$, we plot the maximum amplitude of the noise difference $\Delta T_N$, obtained at $\Vdc=0$, as a function of the corresponding values of $\alpha$ extracted from the fits shown in Fig.~\ref{fig-fig3}. The result is displayed in Fig.~\ref{fig-fig4}: the effect of $\nu$ appears clearly as a deviation from a linear behavior, more pronounced at high frequency. This deviation is well reproduced by a PASN model including heating caused by the THz radiation, shown as continuous lines. In contrast, a time-averaging model, where the noise is given by the average of usual shot noise under a periodic potential, yields a linear variation (dashed lines)~\cite{Reydellet2003}.

We finally analyze the performances of our system as a THz detector. Our data show that despite the low coupling to the emitter, we are able to apply ac voltages across the sample up to 2~mV in the hundreds of GHz range. We also extract the Noise Equivalent Power (NEP), defined as the power detected in a 0.5 s measurement with a signal-to-noise ratio of 1. Our setup allows detecting a typical variation in the noise $\delta T_{N}\left[SNR=1,t_{meas}=0.5~\mathrm{s}\right]\approx 14$~mK. At $\Vdc=0$, this corresponds to values of $\alpha$ between $0.05$ and $0.1$ in the PASN model including heating. When defining the radiation power $P_{\mathrm{ac}}=\Vac^2/(2Z_{rad})$, where $Z_{rad}=376~\Omega$ is the vacuum impedance, we obtain a typical NEP smaller than $10~\mathrm{pW}/\sqrt{\mathrm{Hz}}$. While this value is comparable to the sensitivity of other THz detectors~\cite{Rogalski2011}, it can be largely improved by adapting the sample impedance, and using an optimized microwave-frequencies noise measurement setup~\cite{Parmentier2011}. Note also that heating effects tend to increase the sensitivity (as the noise signal is increased) at the cost of frequency discrimination, as the cusps in the noise at $e\Vdc=h\nu$ characteristic of PASN become smeared. 

In summary, we have observed signatures of PASN in mesoscopic diffusive graphene ribbons, and shown that hallmark out-of-equilibrium phenomena of electronic quantum transport can be extended to energies much larger than the usually probed microwave domain. This allows envisioning fundamental physics experiments where the transport degrees of freedom of a coherent conductor are coupled to the energy spectrum of complex systems, \textit{e. g.} molecules~\cite{Ferguson2002}, as well as the development of new universal THz detectors based on PASN.

\begin{acknowledgements}

We thank N. Kumada, P. Roche, F. Portier and C. Altimiras for fruitful discussions and careful reading of the manuscript, as well as P. Jacques for technical support, M. Westig for help on the THz simulations, and Julian Ledieu from the Institut Jean Lamour for the STM and XPS characterization of our samples. This work was supported by the French ANR (ANR-11-NANO-0004 Metrograph) the ERC (ERC-2008-AdG MEQUANO) and the CEA (Projet phare ZeroPOVA).

\end{acknowledgements}

\cleardoublepage



\section*{Supplementary material}

# Supplemental Material for
# Photon-assisted shot noise in graphene in the Terahertz range

F.D. Parmentier,[1] L.N. Serkovic-Loli,[1] P. Roulleau,[1] and D.C. Glattli[1]

[1]*SPEC, CEA, CNRS, Université Paris-Saclay, CEA Saclay 91191 Gif-sur-Yvette cedex, France*

## EXPERIMENTAL DETAILS

### Samples

The graphene samples were grown using a chemical vapour deposition on copper foils (CVD) technique, under a flow of $H_2$ and $CH_4$. The growth is done under a constant flow of 200 sccm of $H_2$, and a pulsed flow of 2 sccm of $CH_4$ (10 seconds under $CH_4$ flow, followed by 60 seconds without $CH_4$, repeated 300 times for total coverage of the Cu foil). PMMA is spun onto the surface of the foil covered in graphene; after etching the foil in $FeCl_3$, the sheet of PMMA covered graphene is transferred on undoped SiO2 substrates. This recipe consistently yields large scale monolayer graphene sheets, that were characterized with STM, XPS, SEM and Raman spectroscopy, as shown in Fig. S1. During the processing of the devices, Raman spectroscopy was systematically performed to characterize the transferred crystals. The samples were then processed using conventional electron beam lithography. A high temperature annealing was performed on sample C (350 °C during 15 hours), resulting in a 24 % decrease in its two-point resistance.

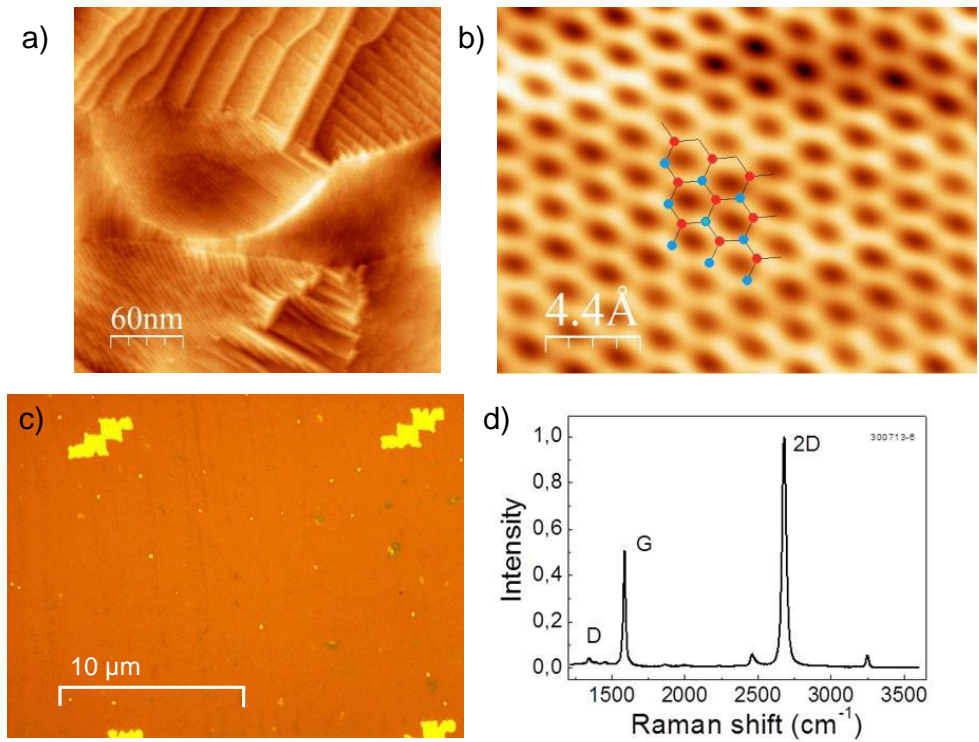

Figure S1. (a) STM image of graphene on copper foil on large scales. Only copper grains are observed due to the dimensions of the image. (b) 2.2×2.2 $nm^2$ STM image of monolayer graphene on copper foil. Carbon atoms forming the honeycomb structure of graphene are observed. (c) SEM image of graphene after transfer onto a SiO2/Si subtrate. (d) Raman spectrum of graphene after transfer onto a SiO2/Si subtrate. The 2D/G peaks ratio is 2.3, indicating that the flakes are monolayer graphene.

The three samples discussed in the article (resp. A, B and C) are monolayer graphene ribbons, respectively (width×length) 1 $\mu$m × 1 $\mu$m, 0.3 $\mu$m × 0.5 $\mu$m, and 1 $\mu$m × 0.6 $\mu$m. The length corresponds to the area of the graphene ribbon which is not in direct contact with the Ti/Au metal electrodes. To decrease the contact resistance, the portion of the ribbon connected to the electrodes is largely increased.

On samples B and C, a pair of side gates is used to modulate the electronic density in the graphene ribbon.

### Measurement setup

The measurements were performed in a dry He3 fridge, with a base temperature of 290 mK. The emitter of the Toptica cw THz generator is thermally anchored to the 2.8 K stage of our fridge. The distance between the emitter and the sample can be changed from a few millimetres to ~ 5 cm.

Simultaneous measurements of low-frequency conductance and noise of the sample are performed using the circuit shown in Fig. S2. The cryoamps used in the noise measurements are home-made amplifiers based Agilent ATF 34143 HEMTs, with gain ~ 4.6 and input voltage noise ~ 0.14 nV/$\sqrt{\text{Hz}}$. The two noise measurement lines allow to perform both auto-correlation noise measurement, as well as cross-correlation noise measurement. Note that in the case of sample B, only a single measurement line was operational. The sample drain-source voltage $V_{\text{dc}}$ is applied through the same lines used for the conductance measurements.

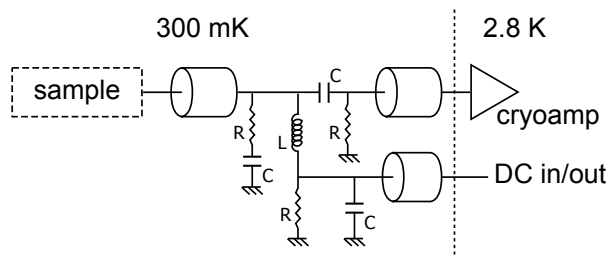

Figure S2. Diagram of the measurement circuit for low-frequency conductance and noise. The values of the different elements in the circuit (corresponding to discrete CMS components) are respectively $R$ = 5 kΩ, $L$ = 22 μH and $C$ = 22 nF. The total shunt capacitance of the *Lakeshore* coaxial wires between the sample and the cryoamp is about 90 pF. A second, identical circuit is connected to the other lead of the sample, allowing for cross-correlation noise measurements.

The noise measurement setup was carefully calibrated on a regular basis to check the gain stability. The calibration was performed by recording the $RLC$ resonance spectra of the two measurement chains at various temperatures between 290 and 600 mK. The various elements of the measurement circuit where calibrated at low temperature in an independent run without sample. In practice, the gains were found to be constant, with fluctuations between different calibrations and cooldowns below 2 %. The noise data obtained from the two auto- and the cross-correlation measurements agree within a few % after calibration.

## ADDITIONAL DATA

### Conductance

Fig. S3 show conductance maps as a function of the gate voltage $V_g$ and $V_{\text{dc}}$ for samples B and C. At low temperature, sample B shows a nanoribbon-like behaviour, with a small energy dependence of the conductance, resulting in unevenly spaced low conductance points at low energy. Sample C shows less effects of the gate voltage, consistent with the fact that the graphene ribbon is significantly wider, decreasing the efficiency of the side gates. The resistance of sample A was ~ 600 Ω. This value provides us with a rough estimate on the contact resistances, of a few hundreds Ω.

At low temperature, the conductance of sample B was found to be affected by the application of THz radiation: in particular, $G_S$ increases at low $V_{\text{dc}}$ as the THz emitter voltage $V_{\text{em}}$ is increased (see Fig. S4). Instead of this being a signature of photon-assisted transport (as, indeed, the conductance is slightly energy-dependent), we believe that it is mainly an effect of heating, which is justified by the fact that the $V_{\text{dc}}$ dependence is small, and when comparing the temperature dependence of the conductance with the data. In particular, the conductance variation between 300 mK and 2.7 K in a similar low conductance point, shown in Fig. S5a), is quite comparable to the variation occurring at zero $V_{\text{dc}}$ in absence of THz excitation, and at maximum THz power (leading to $T_{el}$ = 2.7 K), shown in Fig. S5b).

3

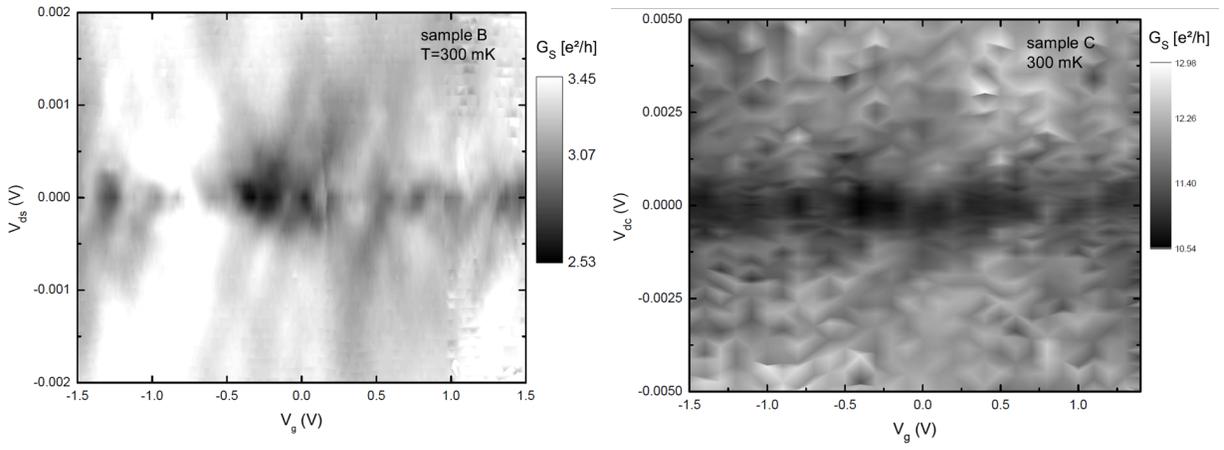

Figure S3. Conductance maps of sample B (left panel) and C (right panel) as a function of the gate voltage $V_g$ and $V_{dc}$.

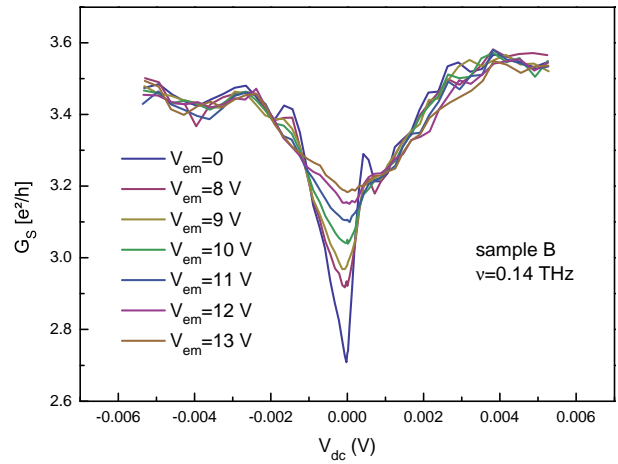

Figure S4. Conductance of sample B as a function of $V_{dc}$, in absence of THz (blue line), and in presence of a 0.14 THz excitation for various $V_{em}$. The colours correspond to the data shown in Fig. 3 of the main article.

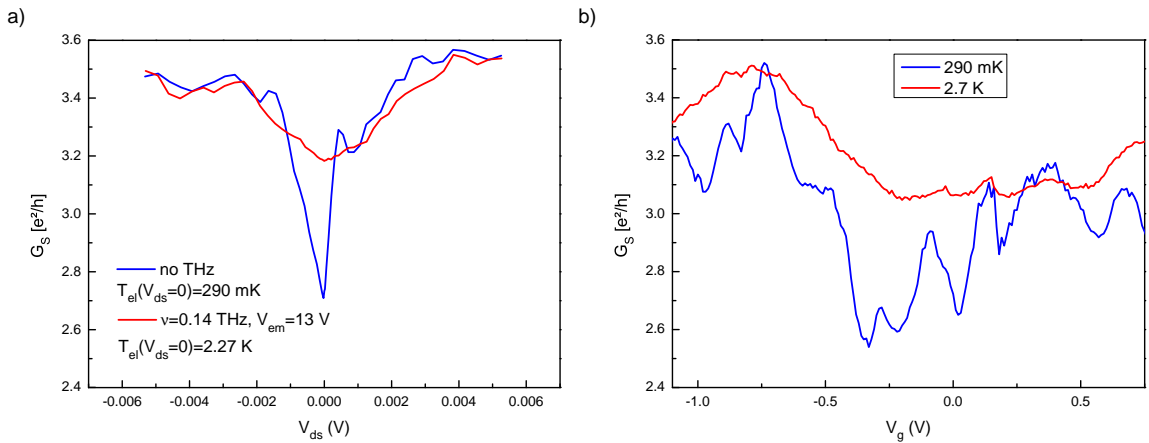

Figure S5. (a) Conductance of sample B as a function of $V_{dc}$, in absence of THz (blue line), and in presence of a 0.14 THz excitation for $V_{em}$ = 13 V, leading to an electronic temperature $T_{el}$ = 2.27 K at zero $V_{dc}$ (red line). (b) Conductance of sample B versus gate voltage $V_g$, around a similar low-conductance point as the one studied in (a), in absence of THz. Blue line: $T_{el}$ = 290 mK, red line: $T_{el}$ = 2.7 K.

## Noise

Fig. S6 shows the dependence of the uncalibrated noise at zero $V_{dc}$ as a function of the THz excitation frequency $\nu$ for sample B (left) and sample C (right). Autocorrelation noise measurements are shown for the former, and cross-correlation measurements for the latter (hence the difference in magnitude). Note that while the data for sample C presents a series of somewhat regularly spaced, broad resonance peaks (similar to the data for sample A shown in the main article, although not as regular), no such feature is observed for sample B (which nonetheless presents clear, sharp resonances). Several other samples (all of them with side gates) were tested, and showed similar sharp, irregularly spaced resonances. This might indicate that the presence of the side gates strongly perturbs the frequency response of the antenna. For all samples, the noise signal vanishes at $\nu > 0.6$ THz.

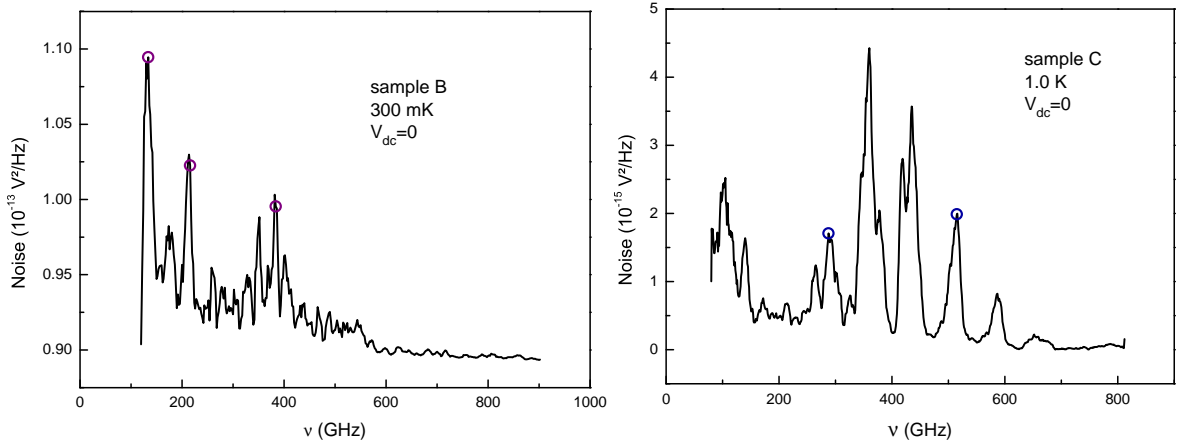

Figure S6. Raw data of the noise measured at zero $V_{dc}$ as a function of the THz radiation frequency $\nu$. Left panel: autocorrelation noise for sample B. Right panel: cross-correlation noise for sample C. The circles represent the frequencies at which the data shown in the main article were taken.

Fig. S7 shows measurement of the difference $\Delta T_N$ between the noise temperature of sample B in presence of a 0.22 THz excitations for various powers ($V_{em}$ ranging from 8 to 13 V) and in absence of THz. The symbols are experimental data, and the lines are results of the model presented in the main article, combining PASN and heating (see also main article Fig. 3 for similar data obtained at 0.14 and 0.39 THz). The fits shown in Fig. S7 are used to extract the data shown in main article Fig. 4.

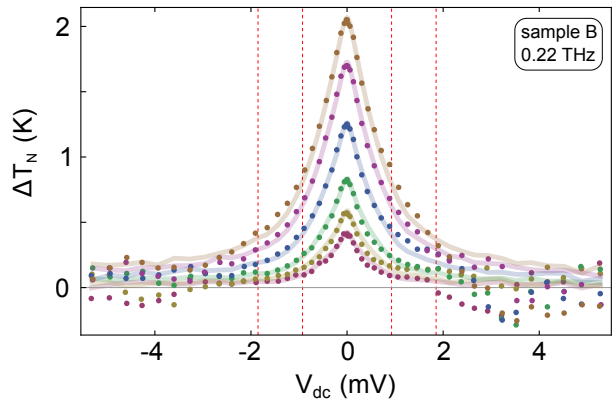

Figure S7. Noise difference $\Delta T_N$ for $\nu = 0.22$ THz, measured at 300 mK on sample B for various $V_{em}$. The symbols are experimental data, and the lines are the fit presented in the main article.

## ANALYSIS

### THz response

In order to gain a better understanding of the frequency response of our devices (see main article Fig. 1(d), and supplementary Fig. S6), we have performed simulations of the frequency response of our design using CST. We have calculated the ac voltage drop $V_{\text{antenna}}$ developing across a bow-tie antenna, under illumination by a THz source. The simulated device has a design close to the lithography pattern used for the samples, without the side gates, but including the inductive blockers used to prevent the ac signal to flow from the antenna to the ground. To simulate the effect of the graphene ribbons, we have included a resistive thin film connecting the two parts of the antenna, the resistance R of which was fixed to R=100 Ω, 1 kΩ and 10 kΩ. To simulate our source, we have used a plane wave with tunable frequency impinging at normal incidence on the sample surface. Note that the specifications of our commercial emitter (which has a log-spiral antenna) indicate that its frequency response is essentially devoid of resonances, but exponentially decreasing.

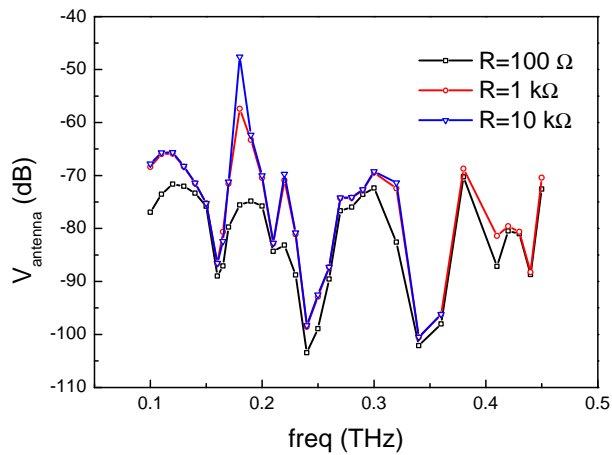

Figure S8. Calculation of the ac voltage drop $V_{\text{antenna}}$ developing between the two pads of a bow-tie antenna, including inductive blockers (see main article Fig. 1), in response to an electromagnetic plane wave with normal incidence with respect to the sample surface. The frequency dependence of $V_{\text{antenna}}$ was calculated for three typical values of the sample resistance: R=100 Ω (black), 1 kΩ (red) and 10 kΩ (blue).

The results are shown in Fig. S8, where we have plotted the frequency dependence of $V_{\text{antenna}}$, expressed in dB with respect to the incident plane wave amplitude. For all three values of R, the calculation shows that the antenna couples rather weakly to the impinging radiation, with $V_{\text{antenna}}$ systematically lower than -50 dB. This is consistent with the very low $V_{\text{ac}}$ extracted from the data shown in the main paper. The signal is markedly smaller for R=100 Ω than for larger values, corresponding to the fact that a low impedance sample will short-circuit the antenna. Furthermore, significant dips regularly appear in the signal (with a spacing of about 100 GHz). This dips do not seem to be affected by the resistance R, which suggests that they stem from resonances due to the antenna geometry. While these results are qualitatively in agreement with the frequency dependence observed experimentally, a quantitative description of the system is hard to obtain, as our simulation neglects the exact frequency dependence of our source, as well as the resonances expected to arise in the cavity formed by the source antenna and the sample antenna. Nonetheless, these calculations show that significant improvements in the high-frequency design of our structures have to be made in order to implement them as potential detectors. These improvements are beyond the scope of this first proof of concept experiment. Finally, we suspect that the side gates might have a strong impact on the frequency response of the bow-tie antenna forming the electrodes of the device. Indeed, only sample A, which doesnt have the side gates, was found to have a somewhat regular frequency response. All other samples had frequency responses qualitatively similar to the ones of sample B (supplementary Fig. S5). Moreover, trying to calculate the response of the antenna with side gates using CST resulted in an error due to a non-converging solution.

**Shot noise analysis**

We now present further details on the analysis of the noise data presented in the main article.

The fit used to reproduce the shot noise data in absence of THz (blue symbols in the upper panel of main article Fig. 2) is based on a simple heat transport model, where the electronic conduction channels in the sample carry the heat dissipated in the sample in presence of a finite bias voltage $V_{dc}$ according to the Wiedemann-Franz law. We thus introduce in the usual shot noise formula:

$$T_N = T_{el} + F\frac{eV_{dc}}{2k_B} \times \left[\coth\left(\frac{eV_{dc}}{2k_B T_{el}}\right) - \frac{2k_B T_{el}}{eV_{dc}}\right] \qquad (1)$$

a bias voltage dependent electronic temperature $T_{el}(V_{dc})$ given by:

$$T_{el}(V_{dc})^2 = T_0^2 + 24\left(\frac{eV_{dc}}{\pi 2 k_B}\right)^2 \times \gamma(1+2\gamma), \qquad (2)$$

where $T_0$ is the base temperature at which the experiment is performed, and $\gamma = G_s/G_c$ is the ratio between the contact conductance and the total conductance of the sample. Using the standard value of the Fano factor $F = 1/3$ for diffusive conductors, we fit the experimental data in absence of THz shown in main article Fig. 2 with $\gamma = 0.0077$, corresponding to $1/G_c \approx 65~\Omega$. Note that using a slightly different value $F = 0.25$ (see below) yields an increased contact resistance $1/G_c \approx 185~\Omega$ still within acceptable bounds.

As the THz emitter output power is not calibrated, the analysis of the noise data $\Delta T_N$ in presence of THz presented in the main article relies on two additional fit parameters, which only depend on $V_{em}$ and $\nu$: the increase in electronic temperature $T_{THz}$ in presence of THz radiation, and the ratio $\alpha = eV_{ac}/h\nu$. The former is inserted in the general expression for the electronic temperature:

$$T_{el}^*(V_{dc}) = \sqrt{T_{el}(V_{dc})^2 + T_{THz}^2} = \sqrt{T_0^2 + T_{THz}^2 + 24\left(\frac{eV_{dc}}{\pi 2 k_B}\right)^2 \times \gamma(1+2\gamma)}, \qquad (3)$$

which is itself inserted into the expression of the PASN given in main article Eq. 1. As explained in the main article, $T_{THz}$ is extracted from the high $V_{dc}$ values of $\Delta T_N$, while $\alpha$ is fixed by the value of $\Delta T_N$ at zero $V_{dc}$, with an electronic temperature thus given by $T_{el}^*(V_{dc} = 0) = \sqrt{T_0^2 + T_{THz}^2}$.

Another possible mechanism to explain the noise increase in presence of THz is time-averaging of shot noise. In this model, the noise is given by the time-averaging of usual shot noise in presence of a periodic potential:

$$T_N = T_{el}^*(V_{dc}) + F\int_0^{1/\nu} dt\nu \frac{eV_{dc} + eV_{ac}\sin(2\pi\nu t)}{2k_B}\left[\coth\left(\frac{eV_{dc}+eV_{ac}\sin(2\pi\nu t)}{2k_B T_{el}^*(V_{dc})}\right) - \frac{2k_B T_{el}^*(V_{dc})}{eV_{dc}+eV_{ac}\sin(2\pi\nu t)}\right]. \qquad (4)$$

Fig. S9 shows the comparison between this model and the PASN model described in the main article, on two sets of data obtained with sample B. Both models include heating due to the THz radiation; $V_{ac}$ is used as a free parameter in the time-averaging model to match the value of $\Delta T_N$ at zero $V_{dc}$. At higher THz frequency ($\nu = 0.39$ THz), the time-averaging model fails to describe the experimental data, particularly at intermediate values of $V_{dc}$; however, both models tend to coincide and reproduce the data at lower frequency $\nu = 0.14$ THz. In general, the two models tend to match for large values of $\alpha = eV_{ac}/h\nu$, corresponding either to small $\nu$ and/or large $V_{ac}$.

This effect appears clearly in main article Fig. 4, where the noise increase $\Delta T_N$ at zero $V_{dc}$ is plotted as a function of $eV_{ac}/h\nu$. To compare our experimental data (shown as symbols in main article Fig. 4) with both models, we extract the dependence of the electronic temperature increase due to the THz radiation $T_{THz}$ with $\alpha$ from the fits performed on $\Delta T_N$. Fig. S10 shows the result of this analysis on the data obtained with sample B at $\nu = 0.14$, $0.22$ and $0.39$ THz, demonstrating a linear dependence for $T_{THz}(\alpha)$ which extrapolates to zero at $\alpha = 0$. We then use this linear dependence to plot the predicted variations of $\Delta T_N(V_{dc} = 0)$ as a function of $\alpha$ for the PASN and the time-averaged models, respectively shown as thick lines and dashed lines in main article Fig. 4. As well as demonstrating the good agreement between our data and the PASN model, this figure shows that the two models indeed coincide at large $\alpha$.

Having extracted the values of $T_{THz}$ and $V_{ac}$ corresponding to each set of $\{V_{em}, \nu\}$, we perform a further test of the consistency of our approach by comparing the extracted $T_{el}^* = \sqrt{T_0^2 + T_{THz}^2}$ and the expected electronic temperature

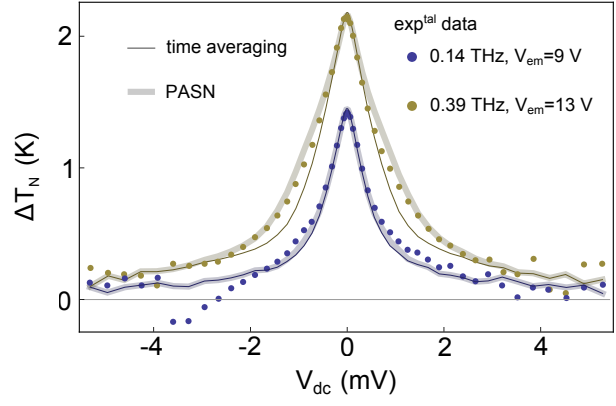

Figure S9. Noise difference $\Delta T_N$ for $\nu$ =0.14 THz and $V_{\rm em}$ = 9 V (blue dots), and for $\nu$ =0.39 THz and $V_{\rm em}$ = 13 V (dark yellow dots), measured at 300 mK on sample B for versus $V_{\rm dc}$. The thick lines are fits using the model combining photon-assisted shot noise and heating, and the thin lines are fits using a model combining time-averaging and heating.

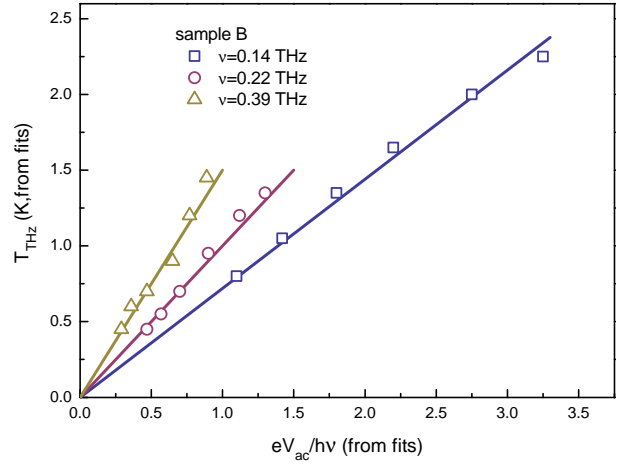

Figure S10. Electron temperature increase $T_{\rm THz}$ in sample B, as a function of $eV_{\rm ac}/h\nu$, for $\nu$ = 0.14 (blue squares), 0.22 (purple circles) and 0.39 THz (dark yellow triangles). Both $T_{\rm THz}$ and $eV_{\rm ac}/h\nu$ are extracted from the fits explained in main article Fig. 2 and shown in main article Fig. 3, and supplementary Fig. S7. The lines are linear fits extrapolating to the origin.

increase $T_{\rm calc}$ due to the finite $V_{\rm ac}$, calculated using the same heat transport formalism and the same parameters as in Eq. 2. Note that this very simple approach completely disregards any capacitive or inductive effects which might be exacerbated, given the frequency range considered here. The result, plotted in Fig. S11, shows that even though both $T_{el}^*$ and $T_{\rm calc}$ have the same order of magnitude, $T_{el}^*$ is in average larger by a factor ∼ 1.5. It is possible that this discrepancy is due to direct absorption of the impinging THz radiation by the graphene flake [22]; because of the very weak electron-phonon coupling in graphene, this absorption can lead to an increase in the electron temperature, that would add to the one calculated above. Note however that using a slightly lower value of the Fano factor $F$ = 0.25 to analyse the noise data in absence of THz (see above) increases the contribution of heating in the noise and yields extracted values of $T_{\rm THz}$ and $\alpha$ larger by a factor 1.12 and 1.06, respectively. Ultimately, the temperatures $T_{el}^*$ and $T_{\rm calc}$ obtained with this value of $F$ are much closer, with within less than 10 % in average. Note that the value $F$ = 0.25 is not unrealistic, as a reduced value of the Fano factor with respect with the expected 1/3 is commonly observed.

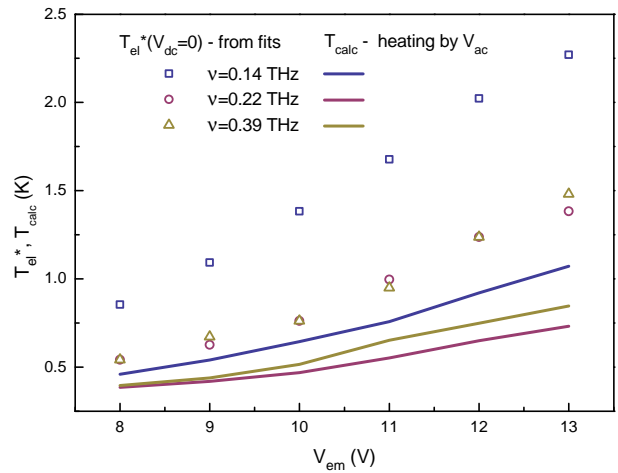

Figure S11. Electron temperature $T_{el}^*$ in sample B, at zero $V_{dc}$, as a function of $V_{em}$, for $\nu$ = 0.14 (blue squares), 0.22 (purple circles) and 0.39 THz (dark yellow triangles). $T_{el}^*$ is extracted from the fits explained in main article Fig. 2 and shown in main article Fig. 3, and supplementary Fig. S7. Lines: $T_{calc}$ calculated with the heating model used in the main article to describe the shot noise data in absence of THz, where the power dissipated in the sample originates from $V_{ac}$ (see text).

---



%

\end{document}